# Ultrafast Spin Dynamics and Photoinduced Insulator-to-Metal Transition in α-RuCl₃

Jin Zhang[1,*], Nicolas Tancogne-Dejean[1], Lede Xian[1], Emil Viñas Boström[1], Martin Claassen[3], Dante M. Kennes[1,4], and Angel Rubio[1,2,5,6*]

[1]*Max Planck Institute for the Structure and Dynamics of Matter and Center for Free-Electron Laser Science, Luruper Chaussee 149, 22761, Hamburg, Germany*

[2]*Center for Computational Quantum Physics (CCQ), The Flatiron Institute, 162 Fifth avenue, New York NY 10010*

[3]*Department of Physics and Astronomy, University of Pennsylvania, Philadelphia, PA 19104, USA*

[4]*Institut für Theorie der Statistischen Physik, RWTH Aachen University and JARA-Fundamentals of FutureInformation Technology, 52056 Aachen, Germany*

[5]*Nano-Bio Spectroscopy Group, Universidad del País Vasco, 20018 San Sebastián, Spain*

[6]*Center for Computational Quantum Physics (CCQ), The Flatiron Institute, 162 Fifth Avenue, New York NY 10010*

*\*Corresponding: Angel Rubio (angel.rubio@mpsd.mpg.de)*

*Jin Zhang (jin.zhang@mpsd.mpg.de)*

## Abstract

**Laser-induced ultrafast demagnetization is a phenomenon of utmost interest and attracts significant attention because it enables potential applications in ultrafast optoelectronics and spintronics. As a spin-orbit coupling assisted magnetic insulator, α-RuCl$_3$ provides an attractive platform to explore the physics of electronic correlations and related unconventional magnetism. Using time-dependent density functional theory, we explore the ultrafast laser-induced dynamics of the electronic and magnetic structures in α-RuCl$_3$. Our study unveils that laser pulses can introduce ultrafast demagnetizations in α-RuCl$_3$, accompanied by an out-of-equilibrium insulator-to-metal transition in a few tens of femtoseconds. The spin response significantly depends on the laser wavelength and polarization on account of the electron correlations, band renormalizations and charge redistributions. These findings provide physical insights into the coupling between the electronic and magnetic degrees of freedom in α-RuCl$_3$ and shed light on suppressing the long-range magnetic orders and reaching a proximate spin liquid phase for two-dimensional magnets on an ultrafast timescale.**

Keywords: ultrafast spin dynamics; insulator-to-metal transition; α-RuCl$_3$; two-dimensional magnets; TDDFT

## INTRODUCTION

In correlated materials, microscopic degrees of freedom (*e.g.*, electrons, phonons, excitons and magnons) are intertwined and their interplay is prominent to understanding the macroscopic properties of quantum materials [1-3]. Photoexcitation with strong laser pulses provides a powerful method to drive correlated materials into out-of-equilibrium states and disentangle the dominant interactions (*e.g.*, electron-electron correlation, spin-orbital coupling and electron-phonon interaction). Upon photoexcitation, fundamental insights can be gained into light-matter interactions, photo-induced phase transitions, and ultrafast dynamics of quasi-particles [4-14].

Photoinduced demagnetization of magnetic insulators paves the way for launching ultrafast dynamics of spins, which cannot be reached in terms of conventional methods to modulate the microscopic magnetism [1,15-16]. Owing to the simple honeycomb crystal structure, the ruthenium-based compound α-$RuCl_3$ provides an attractive platform to explore the physics of electronic correlations, unconventional magnetism and opto-magnetic effects in correlated insulators [17-33]. It is illustrated that α-$RuCl_3$ accommodates essential ingredients of the Kitaev model owing to the interplay of electron correlations and magnetic interactions, facilitating a variety of exotic quantum phases [19-23].

Previous studies have found that α-$RuCl_3$ has a substantial spin-orbit coupling and low-temperature magnetic order, matching the predictions of being a proximate quantum-spin liquid [24]. Recent experiments using angle-resolved photoemission spectroscopy reported a bandgap of ~1.0-1.9 eV, establishing α-RuCl3 as a spin-orbit coupling assisted magnetic insulator [20-21]. Numerous studies have explored the possibility of reaching a phase having gapless spin excitations under magnetic fields and suggested that adequate perturbations are capable of triggering phase transitions among various magnetic phases and correlated states [26-31]. Kasahara *et al*. argued that magnetic fields could destroy the long-range magnetic order and generate a quantum-spin-liquid state or lead to the fractionalization of spins into itinerant Majorana fermions [25]. More relevantly, experiments demonstrated that light pulses can be utilized to tailor the magnetic free-energy landscape of α-RuCl$_3$ and that photoexcitation suffices to induce a quasi-stationary transient spin-disordered phase [30-31]. However, it is elusive whether optical modulations with strong laser pulses can suppress the long-range magnetic order and introduce spin dynamics, which is a promising route to understand correlated physics and calls for studies of the underlying mechanism of photoexcitation under extreme conditions.

In this article, we employ ab initio calculations within the framework of real-time time-dependent density functional theory (TDDFT) [34-35] to investigate laser-driven spin dynamics of the two-dimensional magnet α-RuCl$_3$. To understand its optical response, we undertake a comprehensive evaluation of the electronic and magnetic properties of α-RuCl$_3$, and further simulate its dynamical response to laser pulses with different photon energies and intensities. Based on the recently-developed ACBN0 functional [36-39], the spin dynamics of the correlated insulator as well as ultrafast melting of the bandgap is explored. We find that the photoinduced demagnetization significantly depends on the laser wavelength on account of the photoexcited band renormalization (*i.e,* insulator-to-metal transitions) and carrier excitation. The delicate interplay of the photodoping effect and the insulator-to-metal transition suggests a way to drive the electronic and magnetic structures out of equilibrium on a timescale of tens of femtoseconds.

**RESULTS**

**Electronic and magnetic properties of α-RuCl$_3$ in the ground state**

Figure 1 exhibits the atomic, electronic, and magnetic structures of α-RuCl$_3$. α-RuCl$_3$ is a two-dimensional system with an ideal Ru honeycomb lattice with the Ru-Cl-Ru angle being close to 90°. It hosts comparatively modest spin-orbit coupling in the 4d Ru ions of ~ 0.1 eV [19]. The ground state of α-RuCl$_3$ displays an in-plane zigzag antiferromagnetic (AFM) order, in which the magnetic moments of Ru ions are parallel to other moments in the same zigzag chain and antiparallel to those in neighboring zigzag chains [Figure 1(a)].

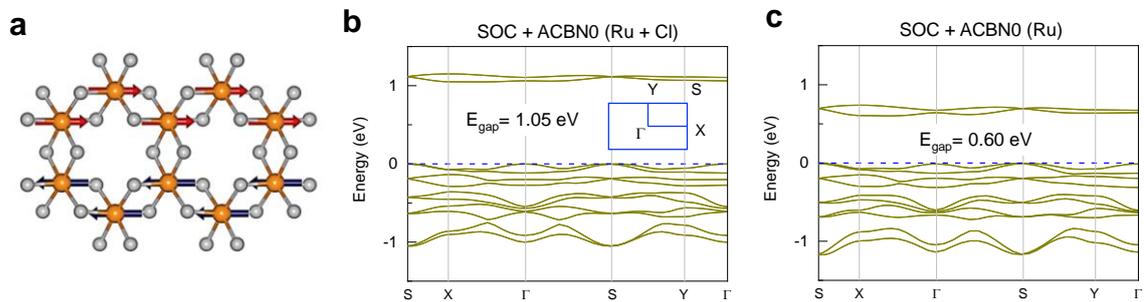

**Figure 1. Atomic, magnetic and electronic structures of α-RuCl$_3$.** (**a**) Atomic structure of α-RuCl$_3$. Orange (gray) spheres denote Ru (Cl) atoms. Red and blue vectors indicate the magnetic moments in the ground state. (**b**) Band structure of α-RuCl$_3$ with on-site Hubbard *U* correction on Ru and Cl orbitals. The effective Hubbard *U* values are respectively 1.96 eV and 5.31 eV for Ru 4d and Cl 3p orbitals after full self-consistency based on ACBN0 functional. Inset shows a schematic of the Brillouin zone with high-symmetry points marked. (**c**) Band structure of α-RuCl$_3$ with on-site Hubbard *U* on Ru 4d orbitals. The effective Hubbard *U* ($U_{eff} = U - J$) is 1.78 eV after full self-consistency. The dashed blue lines (0 eV) indicate the valence band maximum of each panel. Spin-orbital coupling (SOC) is included in all calculations.

We employ the recently proposed ACBN0 functional to exploit the electronic properties of the strongly correlated material [36-39]. The functional is regarded as a pseudo-hybrid reformulation of the density-functional theory plus Hubbard *U* (DFT + *U*) method, enabling us to compute the Hubbard *U* and Hund's *J ab initio* and self-consistently by solving generalized Kohn-Sham equations (See Supplementary Note 1). The method has been recently extended to the real-time case, within the framework of time-dependent density-functional theory. In practice, the functional is an efficient and computationally affordable method to study the optical response of correlated systems driven out of equilibrium [36-39].

As to α-RuCl$_3$, the converged effective Hubbard $U$ terms ($U_{eff} = U - J$) with the ACBN0 functional are 1.96 eV for Ru 4d orbitals and 5.31 eV for Cl 3p orbitals, respectively. The parameters yield an indirect bandgap of E$_{gap}$ =1.05 eV [Figure 1(b)], in excellent agreement with the experimental observation (~1.0-1.9 eV) [20-21]. In contrast, the indirect bandgap of α-RuCl$_3$ with on-site Hubbard $U$ on only Ru 4d orbitals reduces to 0.60 eV [Figure 1(c)], in accordance with previous calculations using empirical Hubbard $U$ terms [32-33]. Furthermore, without on-site Hubbard corrections, the band structure of α-RuCl$_3$ displays a metallic state. These results validate the necessity of on-site terms on both Cl 3p and Ru 4d orbitals.

In previous studies, theoretical calculations based on density functional theory plus a Hubbard correction yield a relatively small bandgap (0.6-0.8 eV) in α-RuCl$_3$ [32-33] while recent experiments reported a larger bandgap of ~1.0 eV [20], indicating an evident inconsistency. With respect to magnetic moments, Banerjee *et al.* [19] reported the ordered moments in Ru$^{3+}$ ions are around 0.4 $\mu_B$ for the low-temperature magnetic phase using neutron diffractions. Based on the ACBN0 functional, the magnetic moments of the Ru atoms are 0.33 $\mu_B$ for the in-plane zigzag AFM order (with Hubbard correction on both orbitals), which is considered to be the ground-state magnetic configuration. These findings reflect that calculations with both spin-orbit coupling and Hubbard $U$ corrections on the Ru 4d and Cl 3p orbitals are able to capture the microscopic interactions and reproduce the experimental electronic structures.

From the projected band structures of α-RuCl$_3$ (see Supplementary Figure S1), the orbitals of both the conduction and valence bands are found to exhibit non-negligible contributions from Cl orbitals, validating that onsite Coulomb potentials on both the Ru 4d and Cl 3p orbitals are critical to obtain accurate electronic and magnetic properties. This is interpreted as the on-site Coulomb potential on the chlorine ions increasing the localization of the lone pairs and hence, the bandgap. Besides, we calculated the amount of charge transfer in the compounds. From Bader analysis (Note 1), the Ru atoms in α-RuCl$_3$ have the charge of 6.94 $e$, in contrast to 8 $e$ in the pristine valence orbitals of the pseudopotential. On the other hand, the charge of each Cl atom is 7.35 $e$ (out of 7 $e$), denoting a significant charge transfer. Therefore, the orbitals of Ru atoms cannot be considered fully localized, and the use of large Hubbard $U$ as a fitting parameter lacks a reasonable physical basis.

Besides the in-plane AFM state, it is observed another possible modulated zigzag antiferromagnet order where the magnetic moments are oriented ±35° from the ab plane in experiments [22]. Our further simulations on α-RuCl$_3$ with the modulated magnetic orders are presented in Supplementary Figure S2, where the magnetic orders are fixed as the starting parameters. For the two magnetic states, the energy difference is very small (10 meV/atom energetically lower for the modulated zigzag state). We observe that α-RuCl$_3$ with the modulated zigzag state exhibits a slightly smaller bandgap of 1.0 eV while the in-plane ferromagnetic state shows a much smaller indirect bandgap of 0.80 eV. The comparison indicates magnetic states are crucial to determine the electronic structures of the system. It should be mentioned that our calculations do not consider the interlayer magnetic interactions because it is hard to resolve the interlayer structure of α-RuCl$_3$ in recent experiments [19]. The weak van der Waals bonding between the α-RuCl$_3$ layers enables several stacking configurations including a rhombohedral phase with space group R3, as well as a C2/m phase. In this regard, the monolayer α-RuCl$_3$ is used as a prototype to investigate the magnetic structure and photoinduced response, as done in other studies [30-31].

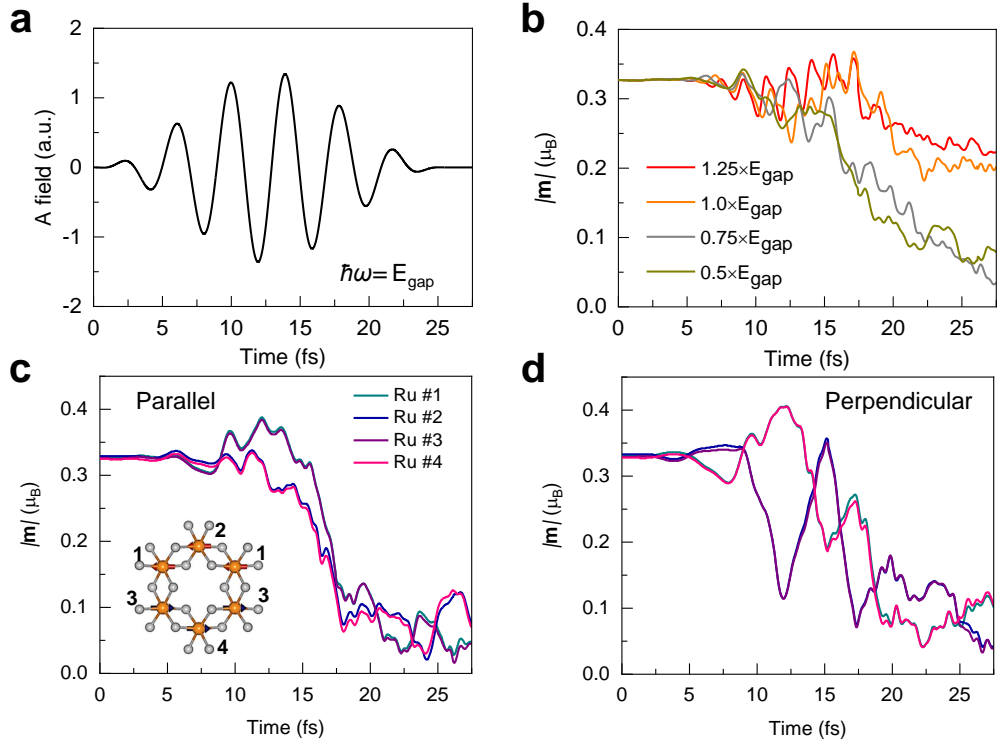

**Figure 2. Laser-induced spin dynamics in α-RuCl₃**. (**a**) Time-dependent vector potential for a wavelength of 1180 nm, corresponding the intensity of $I_0=2.5\times10^{12}$ W/cm². Applied electric fields are in-plane and polarized perpendicularly to the magnetic moment of Ru atoms of α-RuCl₃ with photon energies above or below the ground-state bandgap. The peak value of the laser is at 12.7 fs. (**b**) Time evolution of magnetic moments of Ru atoms under laser excitations with different photon energies ($\hbar\omega=1.25\times$, $1.0\times$, $0.75\times$, and $0.5\times E_{gap}$, respectively). |**m**| indicates the averaged magnet over the values of four Ru atoms in the supercell. (**c**) Dynamics of the magnetic moment of four Ru atoms for laser pulses with parallel polarization and $\hbar\omega=0.5\times E_{gap}$. For in-plane polarization parallel to the Ru moments, Ru #1 follows the dynamics Ru #3, and #2 goes with #4. Inset shows the labels of the Ru atoms. (**d**) The same quantities as shown in (c) for the laser pulses with the polarization perpendicular to the Ru moments. In the perpendicular case, we find that Ru #1 follows the same trend with Ru #4 and Ru #2 with Ru #3. The magnetic moment is calculated by averaging over a sphere of radius 2.22 Bohr around the Ru atoms.

**Wavelength dependence of spin dynamics**

In the following, we focus on the photo-induced out-of-equilibrium dynamics in α-RuCl₃ by altering the laser photon energies. The typical shape of the electric field introduced by the applied laser pulse is shown in Figure 2a, with a wavelength of 1180 nm whose photon energy corresponds to the bandgap (1.05 eV) of α-RuCl₃. Figure 2b summarizes the real-time evolution of the magnetic moments (|**m**|, averaged over the four Ru atoms in the supercell) in α-RuCl₃ under different photon energies above and

below the bandgap.

For a photon energy at the bandgap, a closer inspection of the magnetic moments reveals a clear drop in 25 fs after which they become relatively stable with only a small fluctuation. For lower photon energies, an ultrafast melting of the zigzag AFM magnetic order is observed on a time scale of 20 fs. For a photon energy higher than the bandgap ($\hbar\omega=1.25\times E_{gap}$), the averaged magnetic moments reduce to 0.23 $\mu_B$ and oscillate slightly afterward. We find faster demagnetization processes when considering longer wavelengths corresponding to lower photon energies (Figure 2b). Notably, the residual magnetic moment for $\hbar\omega= 0.75\times E_{gap}$ is roughly 0.04 $\mu_B$ at the end of laser irradiation and similar to the value for $\hbar\omega= 0.5\times E_{gap}$, reflecting a saturation of the ultrafast demagnetization (see Supplementary Figure S3 for snapshots of magnetic moments of α-RuCl$_3$ under laser excitation). The wavelength dependency of spin dynamics provides a novel knob for the high modulation of magnetic states under laser excitation and deserves elaborate investigations.

Given the equilibrium results in Figure. 1, it is clear that the light-induced reduction of *U* is crucial for the observed changes in the magnetic structure. In addition, we observe similar demagnetization for a longer laser pulse of 50 fs, as shown in see Supplementary Figure S4. To understand the above findings, we monitored the effective Hubbard *U* of Ru and Cl orbitals. It is noteworthy that the laser decreases the effective *U* for Ru 4d orbital to 1.50 eV for $\hbar\omega=0.5\times E_{gap}$ (see Supplementary Figure S5). The modification is obviously faster with regard to $\hbar\omega=1.25\times E_{gap}$, in which the residual effective *U* is 1.39 eV. Since the optical excitation is an ingredient of paramount importance to tune the magnetic properties of correlated materials (*e.g.*, charge-transfer insulators [38] and Weyl semimetals [39]), dynamical modification of electron-electron correlations may pave the way to investigate the phase transitions in

α-RuCl$_3$ from a new freedom.

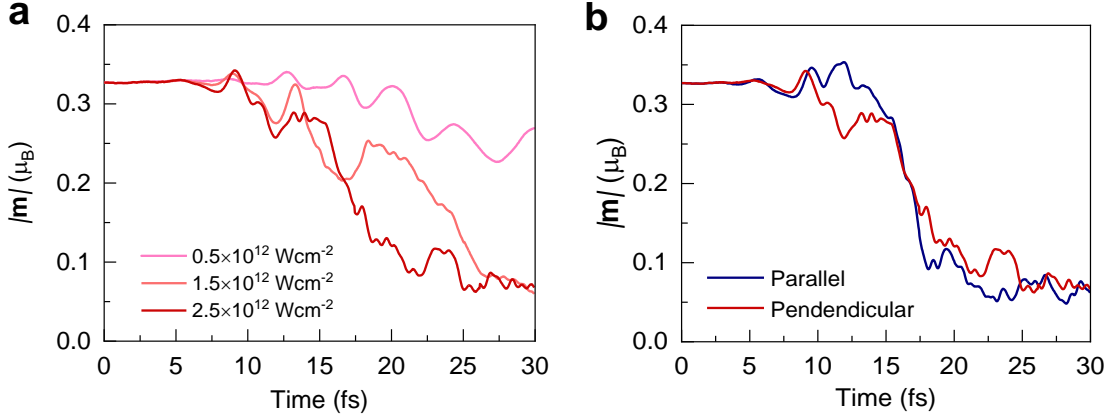

**Figure 3. Intensity and polarization dependence of the light-induced spin dynamics in α-RuCl$_3$.** (**a**) The intensities correspond to I$_0$=0.5×, 1.5×, and 2.5×10$^{12}$ W/cm$^2$, respectively. The laser pulse with perpendicular polarization and higher intensity introduces larger modulation of the atomic magnetic moments. We take the photon energy of $\hbar\omega$=0.5×E$_{gap}$ as an example. (**b**) The influence of laser polarization on spin propagation. Parallel (perpendicular) direction denotes the laser polarization along (perpendicular) to the magnetic moments shown in Figure 1(**a**). The driving intensity and photon energy of laser pulses are I$_0$=2.5×10$^{12}$ W/cm$^2$ and $\hbar\omega$=0.5×E$_{gap}$, respectively.

The variations of the magnetic moments of individual Ru atoms are also provided in Figure 2c-d. For the perpendicular polarization, the photoexcited dynamics demonstrate that the magnitude of spins on different sublattices oscillate with significant out-of-phase components (Panel d and Supplementary Figure S3) and the change follows the frequency of the applied laser pulse. It could be attributed to laser-induced symmetry breaking in charge distributions. We should note that laser pulses with different polarizations are capable of breaking the different symmetries of α-RuCl$_3$, bringing about different spin sublattices in the dynamics. Our calculations reveal that the charge and magnetization dynamics are different for the two sublattices of the honeycomb structure once we apply our laser field, which is attributed to the laser-induced broken symmetries. For the perpendicular polarization, laser excitation breaks the mirror plane vertical to the magnetic moments and we observe two distinct sublattices for demagnetization of Ru orbitals. This is attributed to the symmetry-breaking and the charge redistribution induced by perpendicular excitations (see Supplementary Figure S6).

**Intensity and polarization dependence of spin dynamics**

We also investigate the impact of laser intensity on ultrafast demagnetization. From Figure 3a, it is clear that the laser pulse with a stronger intensity introduces a more considerable modulation of the atomic magnetic moments. For an intensity of $I_0=0.5\times10^{12}$ W/cm$^2$, the residual magnetic moment is 0.21 $\mu_B$, and the zigzag AFM state is still stable after the laser illumination. Whereas for $I_0=2.5\times10^{12}$ W/cm$^2$, we obtain a saturation of the demagnetization at 0.06 $\mu_B$, indicating a stronger reduction and complete melting of the magnetic structure.

Figure 3b illustrates the laser-induced spin dynamics for the laser pulses with polarization parallel to the spins. The spin dynamics follow similar trends as with perpendicular polarization and introduce a distorted state with a residual magnetic moment of 0.04 $\mu_B$, confirming ultrafast demagnetization is robust for the parallel polarization. Our further analysis demonstrates that dynamical modification of the electronic and magnetic parameters in strongly correlated magnets is indeed possible by purely optical means without involving the crystal lattice dynamics. It should be noted that the gap is sensitive to both Hubbard terms and the magnetic orders, indicating the α-RuCl$_3$ is a Mott-Slater insulator. Laser polarization may also be a cardinal ingredient of importance to control the magnetic structures.

**Photoinduced insulator-to-metal transition in α-RuCl$_3$**

We carried out comprehensive calculations for the out-of-equilibrium electronic properties and photoinduced carrier excitations in α-RuCl$_3$. Figure 4a exhibits the transient band dispersion of photoexcited α-RuCl$_3$ for the photon energy of $\hbar\omega=1.25\times E_{gap}$ (see Supplementary Figure S7 for the full trajectory and corresponding bandgaps). The transient band structures and bandgaps are computed from the time-evolved density under various laser excitations, see Supplementary Note 1. The bandgap drops strikingly to 0.24 eV before the spin subsystem responds significantly (in 10 fs). After that, the bandgap melts completely when the laser pulse reaches the peak at about 15 fs, revealing that the band renormalization can take place without any structural distortions in α-RuCl$_3$. We interpret the ultrafast collapse of the bandgap or insulator-to-metal transition in several tens of femtoseconds as indicating that the strong laser pulses greatly change the electron correlations, spin-orbit couplings, and magnetic interactions.

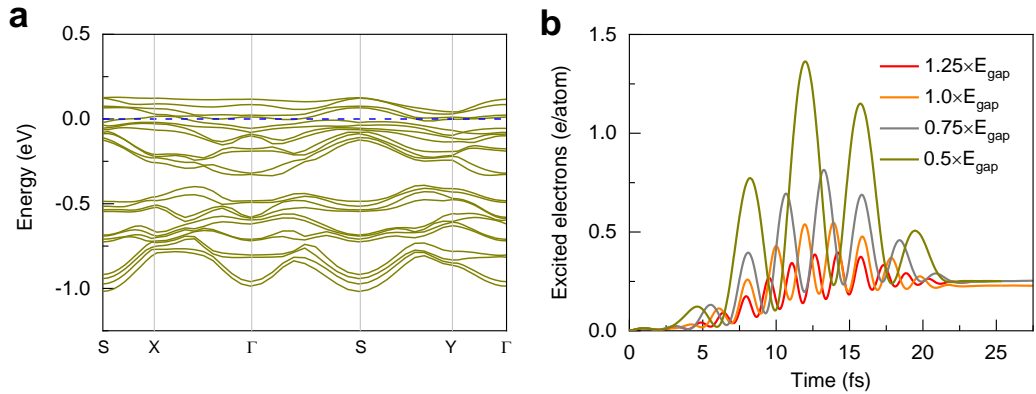

**Figure 4. Photoinduced insulator-to-metal transition and carrier populations for various photon energies.** **(a)** Time-resolved band structures for the laser with the photon energy of $\hbar\omega=0.5\times E_{gap}$ for α-RuCl$_3$ at 25 fs. The dashed blue line represents the Fermi level of the system. **(b)** The photon energies range from $\hbar\omega=0.5\times$ to $1.25\times E_{gap}$. The number of excited electrons is calculated based on the projections of the time-evolved wavefunctions on the ground-state wavefunctions. Here, the intensity is $I_0=2.5\times10^{12}$ W/cm$^2$ and the polarization is perpendicular to the magnetic moments. The results indicate that smaller photon energies lead to higher excited carrier densities owing to the collapse of the bandgap.

Furthermore, the ultrafast insulator-to-metal transition is robust for various photon energies at the laser with strong intensity (see Supplementary Figure S8). Regarding all photon energies, the ultrafast band renormalization takes place within 15 fs. This explains the smaller modulation of magnetic moments. Therefore, the bandgap of α-RuCl$_3$ can be easily modulated by optical excitation. As a direct consequence, the required excitation energy for Zener tunneling or multiphoton ionization decreases during the laser irradiation in α-RuCl$_3$.

In order to elucidate the origin of the optical response, the excited carrier densities are analyzed by characterizing the charge excitation from the valence to conduction bands of Kohn-Sham orbitals, as displayed in Figure 4b. Under different laser intensities, most of the excited electrons are virtually excited, which means the value of excited electrons oscillates following the shape of laser fields and returns almost to zero after each half cycle. We note the oscillation of the population with the field is an artifact of the gauge field projection and not a feature of the time-dependent populations. At the end of the laser pulse (~25 fs), a small amount of the valence electrons is effectively excited into the conduction bands, which play a role in the magnetic dynamics. For $\hbar\omega=1.25\times E_{gap}$, the photoinduced carrier population increases to 0.40 $e$/atom within 15 fs and then oscillates around 0.25 $e$/atom. Following the shapes of the laser pulses, the peak of carrier densities reaches 1.36 $e$/atom for $\hbar\omega=0.5\times E_{gap}$. The excited carrier concentration is also sensitive to the photon energies of the laser, *i.e.*, longer wavelengths result in more significant carrier densities. This is attributed to the laser-induced collapse of the bandgaps and the laser pulses with smaller photon energies becoming resonant with the transient bandgaps.

To validate the physical picture, we performed additional simulations from the modulated zigzag antiferromagnet order to track the magnetic dynamics and transient band structures (see Supplementary Figure S9). It is obvious that the laser-induced magnetic and electronic dynamics are similar for the two possible magnetic states (in-plane zigzag and modulated zigzag orders). Therefore, we obtain a robust picture of ultrafast photoinduced insulator-to-metal transition in α-RuCl$_3$. The collapse of the bandgap occurs when the electrons are excited by strong laser pulses. The saturation at the half of the ground-state bandgap is interpreted as the excited carrier density being high enough to modulate the electronic structures, transient bandgaps and introduce the rapid demagnetization. In addition, non-linear excitation processes can also play a role in the wavelength dependence of the demagnetization, especially at the beginning of the strong laser pulses. Our findings support that the photoinduced carriers are important to introduce ultrafast melting of magnetic structures. Notably, the recovery process after demagnetization is not traced in this work because the real-time TDDFT method incorporates no effective energy dissipation channel. We cannot describe the processes in the time scale of tens of femtoseconds due to a structural change and effective spin-lattice scattering. Moreover, on shorter timescales, electron-electron scattering is expected to play some role, but the scattering channel is out of reach for the currently existing functionals, because of the underlying adiabatic approximation employed in the simulations [34-35].

The direct electronic and spin dynamics obtained from our first-principles calculations enable us to simulate the photoinduced response of α-RuCl$_3$ at the atomistic spatial scale on a femtosecond timescale. Strong photoexcitation leads to an ultrafast insulator-to-metal transition and creates a high density of electron-hole pairs. The magnetic interactions are modulated as an effective non-magnetic state is obtained. It should be noted that the ultrafast process is not the result of collective magnetic excitation (e.g., magnons), which would keep the magnitude of the magnetic moment fixed, while decreasing its components along the local order parameter directions. We note that extracting information about the collective excitation should demand more effort from the model Hamiltonian and TDDFT methods [34-35]. In turn, the melting of the magnetic structure and reduction of the effective Hubbard terms contribute to the decrease of electronic bandgaps since the electronic correlations are crucial to determine the Mott gap.

In α-RuCl$_3$, the collapse of the bandgap occurs when the electrons are excited by strong laser pulses. Our findings support that the photoinduced carriers are important to induce ultrafast melting of magnetic structures. Our results indicate that both the magnetic ordering and the Hubbard *U* are crucial to stabilize the insulating state. Thus, the melting of the bandgap is caused by a conspiracy of the dynamically reduced Hubbard terms and the melting of the magnetic moments. It is straightforward to understand that high power can easily destroy the spin orders in an ultrafast timescale because high-power laser pulse can induce higher carrier density. As for the wavelength dependence, the laser pulse with photo energy around half of the bandgap becomes more resonant with the transient bandgap and excite more carriers, which destabilize the magnetic orders. In experiments, the magnetic transition could be probed by the time-resolved linear magnetic dichroism and x-ray magnetic circular dichroism [32-33], which is able to illustrate the magnetic difference. Photoinduced insulator-to-metal transition in α-RuCl$_3$ can be detected in time- and angle-resolved photoemission spectroscopy. It is noteworthy that a femtosecond laser-induced quantum-spin-liquid state is beyond the current study and will be the subject of future studies.

**DISCUSSION**

In conclusion, our *ab initio* simulations revealed the nature of photon-driven electron and spin dynamics in α-RuCl$_3$. We demonstrated that laser pulses can provoke a magnetic transformation between zigzag AFM magnetic order and disordered magnetic states with much smaller magnetic moments. In addition, the spin response is remarkably sensitive to the laser wavelength and polarization on account of the photoinduced insulator-to-metal transition and different excited carrier distributions. This subtle interplay suggests a way to modulate the electronic and magnetic structures using ultrashort laser pulses. Our work provides new insights into photoexcitation-induced magnetic phase transitions and may pave the way for suppressing the long-range magnetic order and realizing a quantum-spin-liquid state at ultrashort timescales.

**METHODS**

**TDDFT simulations.** The evolution of the spinor states and the evaluation of the time-dependent Hubbard U and magnetization are computed by propagating the generalized Kohn–Sham equations within time-dependent density functional theory including mean-field interactions, as provided by the Octopus package [34-35], using the ACBN0 functional together with the local density approximation (LDA) functional for describing the semi-local DFT part [36-37]. We compute ab initio the Hubbard U and Hund's J for the 4d orbitals of Ruthenium and 3p orbitals of Chlorine. In the time-dependent simulations, the laser is coupled to the electronic degrees of freedom via the standard minimal coupling prescription using a time-dependent, spatially-homogeneous vector potential A(t), with the electric field $E(t) = -\frac{1}{c}\frac{\partial A(t)}{\partial t}$, where c is the velocity of light in vacuum. We consider a laser pulse of 12.7 fs duration at full-width half maximum with a sin-square envelope corresponding to a total width of 25.4 fs. In all our calculations, a carrier-envelope phase of ϕ = 0 is used.

The experimental lattice parameters (i.e., 5.98 Å and 10.35 Å) and atomic positions are employed [19]. We employed a mixed periodic boundary condition with a vacuum region of 15 Å to ensure that interactions between periodic images were negligible. We employ norm-conserving HGH pseudopotentials [40], a real-space grid spacing of 0.33 atomic units, and an 8 × 6 × 1 k-point grid in a 1×√3 supercell with a rhombus shape (containing 16 atoms for the zigzag AFM magnetic order). The inclusion of semi-core states of Ru and Cl elements are prominent to obtain accurate electronic structures; the valence electrons explicitly included are Ru: $4s^2$, $4p^6$, $4d^7$ and $5s^1$; Cl: $3s^2$ and $3p^5$. In all calculations, we include the spin-orbit coupling, which is vital to obtain the correct electronic and magnetic structures.

## DATA AVAILABILITY

The data that support the findings of this study are available from the corresponding author upon reasonable request.

## ACKNOWLEDGEMENTS

This work is supported by the European Research Council (ERC-2015-AdG-694097), Grupos Consolidados (IT1249-19), and SFB925. AR is supported by the Flatiron Institute, a division of the Simons Foundation. We acknowledge funding by the Deutsche Forschungs-14 gemeinschaft (DFG, German Research Foundation) under RTG 1995, within the Priority Program SPP 2244 "2DMP", under Germany's Excellence Strategy - Cluster of Excellence and Advanced Imaging of Matter (AIM) EXC 2056 - 390715994 and RTG 2247. JZ acknowledges funding received funding from the European Union's Horizon 2020 research and innovation program under the Marie Sklodowska-Curie grant agreement No. 886291 (PeSD-NeSL). LX acknowledges the support from Distinguished Junior Fellowship program by the South Bay Interdisciplinary Science Center in the Songshan Lake Materials Laboratory. JZ thanks Michael Sentef for fruitful discussions.

## AUTHOR CONTRIBUTIONS

AR conceived the research direction and JZ proposed the project. Most of the calculations were performed by JZ, with contributions from all authors. NTD implemented the ACBN0 functional into the OCTOPUS code. All authors contributed to the analysis and discussion of the data and the writing of the manuscript.

## COMPETING INTERESTS

The authors declare no competing interests.

## ADDITIONAL INFORMATION

Correspondence and requests for materials should be addressed to J.Z. or A.R.

**TOC figure**

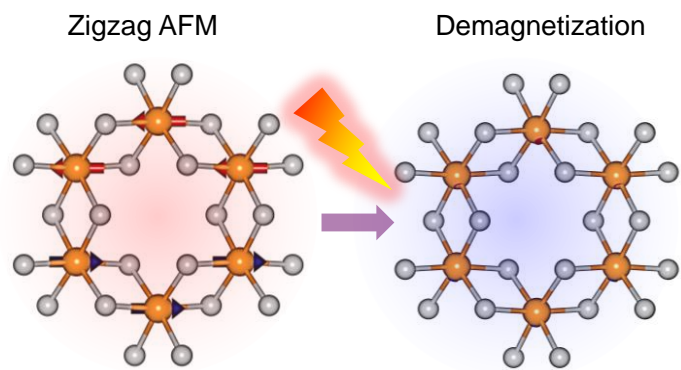